\documentclass[a4paper,aps,pre,superscriptaddress,floatfix,nofootinbib,twocolumn,showpacs,bibtex]{revtex4}

\usepackage{graphicx}
\usepackage{latexsym,amsmath,verbatim}
\usepackage{color}
\usepackage[english]{babel}

\begin{document}

%\title{An evolutionary model explains the emergence of optimal L\'evy-flight strategies in mental searches}
\title{Evolution of optimal L\'evy-flight strategies in human mental searches}

\author{Filippo Radicchi}\email{f.radicchi@gmail.com}
\affiliation{Departament d'Enginyeria Quimica, Universitat Rovira i Virgili, Av. Paisos Catalans 26, 43007 Tarragona, Catalunya, Spain}

\author{Andrea Baronchelli}\email{a.baronchelli.work@gmail.com}
\affiliation{Laboratory for the Modeling of Biological and Socio-technical Systems, Northeastern University, Boston MA 02115 USA}

\date{\today}

\begin{abstract}
\noindent Recent analysis of empirical data [F. Radicchi, A. Baronchelli \& L.A.N. Amaral. PloS ONE {\bf 7}, e029910 (2012)] showed that humans adopt L\'evy flight strategies when exploring the bid space in on-line auctions. A game theoretical model proved that the observed L\'evy exponents are nearly optimal, being close to the exponent value that guarantees the maximal economical return to players. Here, we rationalize these findings by adopting an evolutionary perspective. We show that a simple evolutionary process is able to account for the empirical measurements with the only assumption that the reproductive fitness of the players is proportional to their search ability. Contrarily to previous modeling, our approach describes the emergence of the observed exponent without resorting to any strong assumptions on the initial searching strategies. Our results generalize earlier research, and open novel questions in cognitive, behavioral and evolutionary sciences.
\end{abstract}


\maketitle

\section{Introduction}
\noindent L\'evy flights
are a special class of random walk whose step lengths
follow a power-law tailed distribution~\cite{shlesinger93}. 
They have been proved to be the most efficient type of space exploration that can 
be adopted by a random searcher looking for scarce
resources in an unknown environment~\cite{viswanathan99}.
Probably for this reason, there are plenty of 
empirical evidences that movement patterns are compatible with L\'evy
flights in 
many different contexts where efficiency
matters~\cite{mantegna95, viswanathan96, ramos04, bartumeus03, bertrand05, 
brockmann06, rhodes07, sims08, humpries10, viswanathan10}. 
In particular, L\'evy flights appear recurrently
in the description of the motion 
of animals in real space (see~\cite{viswanathan08} for a review).
Animals explore the environment mainly
for searching food resources, and it is therefore 
plausible to ascribe the optimality
of their search strategies to a selective evolutionary
process.

\

\noindent Recently, Radicchi {\it et al.} have provided
empirical evidence that also human players
participating in on-line auctions explore the bid
space performing L\'evy flights~\cite{radicchi11}. 
The exploration of the
bid space represents a search process, but 
purely mental because performed 
in an abstract space. Interestingly, players adopt
nearly optimal L\'evy flight exponents, in the
sense that the values of the exponent used
in real auctions are close to the one that
maximizes their economic return.
In~\cite{radicchi11}, the search process
in the bid space is studied as
a game theoretical model,
where the optimal exponent value corresponds
to a Nash equilibrium~\cite{nash50}.

\

\noindent Here, 
%we study the same problem 
%but from a different
%perspective. 
we propose an evolutionary 
model in which the reproductive fitness of the individuals 
is proportional to their ability to win the
auctions, and we show that the values of
the L\'evy flight exponents to which the model converges
are very close to those measured in real data. 
This approach relaxes some of the assumptions made in the
traditional game theoretical analysis, and deepens
the understanding of the
results: The optimality of the strategies adopted
by bidders in on-line auctions can be seen as the outcome
of a (evolutionary) learning process.

\

\noindent The paper is organized as follows. In section~II,
we provide a detailed description of the type of
auctions studied and modeled in this paper. Section~III
is devoted to the description of the model
and its analytical treatment. In particular,
section~IIIA describes the case in which all players
can choose only a bid value, while, in section~IIIB, we
generalize the model to the case in which players can
place an arbitrary number of bids. Section~IIIC is dedicated
to the evolutionary game theoretical implementation
of the model. 
Sections~IV and~V are respectively devoted to the description
of the numerical simulations of the model and to
the estimation of the computational complexity needed to simulate
or solve the model. In section~VI, we provide a detailed description
of the results of the model. Finally, in section~VII, 
we draw our final comments and considerations.

\section{Lowest unique bid auctions}
\noindent Lowest Unique Bid (LUB) auctions are
a recent generation of online games where the winners of the auctions
may purchase expensive goods for strikingly small
prices: cars, boats and even houses
can be bought for only tens or hundreds
of dollars (or euros, pounds, etc.).
The mechanism of the game is very simple.
At the beginning of the auction, 
a good, whose typical market price is higher than
a thousand dollars, is put up for auction. The game duration
is {\it a priori} fixed and generally two or three
weeks long. A bid can be
any amount (in cents) from one cent to a certain
maximum value $M$, generally lower than one hundred dollars. 
For placing a bid, a player
has to pay a fee, whose entity
ranges from one to ten dollars depending
on the game.
During the auction,
players know only 
the status of their own bids meaning
whether they are winning or not.
None of the players knows where the others have placed
their bids until
the end of the game.
Multiple bids on the same value
from the same player are allowed, but
do not influence the outcome of the auction
since a bid is considered as unique when a
unique player has bid that value even if more
than once.
When the time dedicated to the auction expires,
the winner is the player who made the LUB
and can finally purchase the good
for the value of the winning bid.
For example, if at the end
of the auction bid value $i=1$ is occupied
by two bids ($n_1=2$), while $n_2=3$, $n_3=1$, $n_4=2$
and $n_5=1$ are the number of bids placed on values 
$i=2$, $3$, $4$ and $5$, respectively, then the winner
is the player who has bid on value $i=3$ because
this is the lowest bid among all unique bid values.
\\
A similar mechanism is also used in 
Highest Unique Bid (HUB) auctions.
In this type of the games, the rules are the same as those
of LUB auctions, with the only difference that the winning
value is the unmatched bid closest to the maximal
bid value admitted in the game (i.e., $M$).
\\
LUB and HUB auctions represent
competitive arenas where players  perform
searches for a single target whose position
is determined by the bids of the whole population.
It is important to stress that, during the game, players are not
aware of the values on which the other players
have placed their bids, and therefore
the exploration of the bid space of each player
can be considered independent. Also, since the cost of each bid
is much larger than the natural unit of the game (one cent),
the number of steps that can be performed by a single player
is limited and allows only a partial exploration of
the bid space. Players need therefore good strategies 
in order to maximize their winning chances and simultaneously
maintain limited their investments.

%\

%We collected data from different web sites organizing
%this kind of auctions: {\tt www.uniquebidhomes.com} and
%{\tt www.lowbids.com.au} organizing LUB auctions, and
%{\tt www.bidmadness.com.au} specialized in the organization
%of HUB auctions. Our data sets contain the whole information
%about the games played, including
%the details of each bid: its value, when it was made and
%who did it. These data allow to
%keep track of all the movements performed in the bid space 
%by single players in single auctions. Data are available
%for downolad also at {\tt filrad.homelinux.org/resources}.
%%%%%%%%%%%%%%%%%%%%%%%%%%%%%%%%%
%%%%%%%%%%%%%%%%%%%%%%%%%%%%%%%%%

\

%%%%%%%%%%%%%%%%%%%%%%%%%%%%%%%%%
%%Results of the empirical analysis
%%%%%%%%%%%%%%%%%
\begin{figure}[!htb]
\begin{center}
\includegraphics[width=0.45\textwidth]{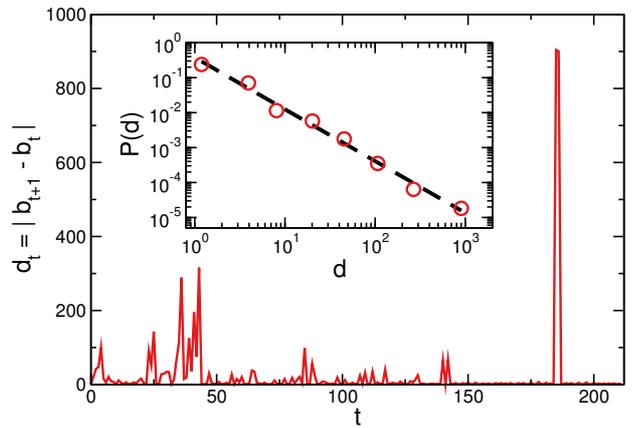}
\caption{(Color online) In the main panel, we plot $d_t$ the absolute
value of the difference between the two consecutive bid vales
$b_{t+1}$ and $b_t$. $t$ indicates the rank position
of each bid in a ranking in which
bids have been sorted accordingly to their time stamp. 
In the inset, we plot the pdf $P\left(d\right)$ of the bid 
change amounts $d$ (red circles).
The pdf is fitted with a power-law function and the
best estimate of the decay exponent is $1.5\pm0.1$ (black dashed line). 
This figure refers to the exploration of the bid space
performed by a single player in a single auction. A complete
analysis of the movement patterns of hundreds of players 
was performed in our previous work~\cite{radicchi11}.
}
\label{fig:example}
\end{center}
\end{figure}
%%%%%%%%%%%%%%%%%
\noindent In our previous work~\cite{radicchi11}, we have studied in detail
the dynamical features of the bid space exploration
performed by players in real LUB and HUB auctions. 
We have found that the exploration of the bid space is bursty:
consecutive bid values are generally close to each other,
but from time to time players perform longer jumps.
In particular, the probability
density function (pdf) $P\left(d\right)$ of the bid change
amount $d$ ($d$ is defined as the absolute value of the difference
between two consecutive bid values) is consistent 
with a power-law $P\left(d\right) \sim d^{-\alpha}$ (see Fig.~\ref{fig:example}).
The exploration of the bid space is therefore consistent
with a discrete version of a L\'evy flight~\cite{hughes81}.
More importantly, we have found that the pdf $g\left(\alpha\right)$
of the L\'evy flight exponents, adopted in real
auctions, is peaked around the
average value $\langle \alpha \rangle \simeq 1.4$
and with standard deviation equal to $\sigma \simeq 0.2$. 
\\ 
The empirical observation of L\'evy flights provided
in~\cite{radicchi11} is by far the most
significant evidence of this phenomenology in
natural search processes.  
Differently from previous studies regarding
biological~\cite{viswanathan99, viswanathan96, bartumeus03, ramos04,  sims08, humpries10} 
and mobility~\cite{bertrand05, brockmann06} 
systems  where 
``two orders of magnitude of scaling can represent a luxury''~\cite{viswanathan08}
, the power-law decay can be clearly observed even over four orders
of magnitude.
The reason is that the space is not
strictly physical and
movements of tens of thousands cents can be performed
at the same price of those one cent long: players
explore the bid space in a effectively {\it super-diffusive} fashion,
and steps are made at a virtually {\it infinite} velocity.

\section{Modeling  bidding strategies}
\noindent In this section, we provide
an analytical model for LUB auctions, but
the model can be easily extended also to HUB auctions.
Supported by empirical evidence,
in our model we assume that players  explore
the bid space performing L\'evy flights.
In our previous work~\cite{radicchi11}, we have provided
a  model based on stronger assumptions.
In particular, we have studied the winning chances of a player
participating in an auction against a population of players using exactly
the same exponent value. Here differently, we do not make any assumption
regarding the choice of the exponents. We let
the exponents to be random variates extracted from a
generic pdf, and
study an evolutionary theoretical version 
of the game. Independently of the initial pdf, the pdf 
of the adopted exponents naturally evolves to a stable distribution
centered around a value remarkably close to the one
of the exponents measured in real auctions.

\subsection{First bid}

\noindent We consider a population of $N$ players whose
strategies $\alpha$ are randomly extracted from a
pdf $g\left(\alpha\right)$.
Without loss of generality, we assume that $g\left(\alpha\right) \geq 0$
if $\alpha \in \left(\alpha_1, \alpha_2\right)$, and $g\left(\alpha\right) =0$
otherwise.
During the game, each player can place bids only on
integer values $i \in \left[1,M\right]$. Since,
at the beginning of the game, none of the players knows 
on which values the others will bid, it is 
natural to think that a generic player, with strategy $\alpha$,
sits at the leftmost site of the lattice.
From this initial positions, the player places a bid
on value $i$ with probability
\begin{equation}
s\left(i, \alpha\right) = \frac{i^{-\alpha}}{m\left(\alpha\right)} \; g\left(\alpha\right) \;,
\label{eq:single} 
\end{equation}
where $m\left(\alpha\right) = \sum_{i=1}^M\, i^{-\alpha}$ is the
proper normalization constant. The probability that 
a generic player bids on value $i$ can 
be calculated
by simply integrating Eq.~(\ref{eq:single}) as
\begin{equation}
p \left(i\right) = \int_{\alpha_1}^{\alpha_2}  d\alpha \; s\left(i, \alpha\right) \;\; .
\label{eq:single_gen} 
\end{equation}
After all players have bid, there will be $n_k$ bids on the $k$-th bid value.
The probability
to observe a particular configuration $\left\{ n \right\} = \left(n_1, n_2, \ldots, n_k, \ldots , n_M\right)$
is simply given by the  multinomial distribution
\begin{equation}
P \left( \left\{ n \right\} \right) =  N! \; \prod_{k=1}^M \, \frac{\left[p \left(k\right)\right]^{n_k}}{n_k!} \;\;,
\label{eq:prob_configuraton} 
\end{equation}
whose weights given by Eq.~(\ref{eq:single_gen}) and
obeying the constraint $ N = \sum_{k=1}^M \, n_k$.
In particular, the probability that only one bid
has been made on value $i$ (i.e., the bid on value $i$ is unique) is 
%\begin{widetext}
\begin{equation}
u \left(i\right) %  = P \left( n_i=1 \right) 
=  \sum_{\sum_{k \neq i} n_k = N-1} P \left( \left\{ n \right\} \right) =  N p \left(i\right) \left[ 1 - p \left(i\right) \right]^{N-1}  . 
\label{eq:unic}
\end{equation}
%\end{widetext}
The probability that the bid on value $i$ is the unique and lowest bid
can be exactly calculated by summing the probability of Eq.~(\ref{eq:prob_configuraton})
over all configurations $\left\{ n \right\}$ which satisfy this constraint
(i.e., $n_i=1$ and $n_j \neq 1$ for all $j<i$). Unfortunately,
such enumeration cannot be easily computed. 
A good approximation, valid for 
sufficiently low values of $p\left(i\right)$
and $u\left(i\right)$,
is to consider the uniqueness of the $i$-th bid value as independent of the 
uniqueness of the other bid values and write
\begin{equation}
l \left(i\right) = \left\{
\begin{array}{l l}
u\left(i\right) & \textrm{ , if } i=1\\
 u \left(i\right) \;  \prod_{k<i} \, \left[ 1 - u \left(k\right)\right] & \textrm{ , otherwise} 
\end{array}
\right.
\;\; ,
\label{eq:unic_low} 
\end{equation}
as the probability that the
bid on value $i$ is the lowest bid among all the unique bids.
The r.h.s. of Eq.~(\ref{eq:unic_low}) is the product of 
two terms: $u \left(i\right)$
 is the probability that only one bid has been made on bid value $i$; 
$\prod_{k<i} \, \left[ 1 - u \left(k\right)\right]$ is the 
probability that none of the bid values smaller than $i$ are
occupied by a single bid (if $i=1$ this probability
is automatically equal to one since $i=1$ is
the minimal bid amount allowed). Finally, 
the probability $w \left(\alpha\right)$ 
that the L\'evy flight exponent $\alpha$ is 
the winning strategy in the game can
be inferred with
\begin{equation}
w \left(\alpha\right) = \sum_{i=1}^M \;  v\left(\alpha \left| i \right. \right) \; l \left(i\right) = \sum_{i=1}^M \; \frac{ s\left(i, \alpha\right) \; l \left(i\right)} {p \left(i\right)} \;\; .
\label{eq:prob_win}
\end{equation}
$v\left(\alpha \left| i \right. \right) = s\left(i, \alpha\right) / p \left(i\right)$ is the conditional probability
to observe $\alpha$ given $i$. This quantity is 
then convoluted over all bid values $i$, where 
the weight of each bid value is given by the probability
that $i$ is the winning bid value, that is $l \left(i\right)$. 
Notice that Eqs.~(\ref{eq:unic}),~(\ref{eq:unic_low}) and~(\ref{eq:prob_win})
explicitly depend on the number of 
players $N$ and the upper bid value $M$. 
We have suppressed both variables
in the notation only for clarity and shortness.

\subsection{Multiple bids}
\noindent The same theoretical approach can be applied
for the determination of the best strategy in a game
where every player may perform multiple bids. 
We consider
the simplest case in which each player bids $T$ times, but
the theory may be simply extended also to the  case in which the number
of bids of each player is extracted from an arbitrary pdf.
\\
In order to solve this game, we need to calculate
$s_T\left(i\right)$ which stands for the probability
that, in $T$ bids, a generic player has placed a bid on value $i$.
If $\alpha$ is the player's strategy, the player will place
the first bid on value $i$ with probability 
$q_1\left(i | \alpha \right) = i^{-\alpha}/m\left(\alpha\right)$. 
For the subsequent bids, we need to define a transition matrix $Q_\alpha$.
The generic element $\left(Q_\alpha\right)_{ji}$ represents the probability
to place a bid on value $i$ when the previous bid was placed on value $j$. 
In our model, we have
\begin{equation}
\left(Q_\alpha \right)_{ji} = \frac{\left| i - j \right|^{-\alpha}}{m_j\left(\alpha\right)} \;\; ,
\label{eq:transition}
\end{equation} 
for all $i$ and $j$ in the interval $[1,M]$. The normalization constant
$m_j\left( \alpha \right) = \sum_{i=1}^M \left| i - j \right|^{-\alpha} $ 
ensures the proper definition of the transition matrix. The matrix $Q_\alpha$
describes a random walker which follows 
uncorrelated L\'evy flights with exponent $\alpha$.
At a generic step $t$, the probability that the player with strategy
$\alpha$ sits on value $i$ is
\begin{equation} 
q_t \left( i \left| \alpha \right. \right) = \sum_{j=1}^M \; \left( Q_\alpha \right)_{ji} \; q_{t-1} \left( j \left| \alpha \right. \right) \;\; .
\label{eq:ran_walker} 
\end{equation} 
The probability that this player placed a bid, in $T$ bids,
on value $i$ is then
\begin{equation} 
s_T\left( i \left| \alpha \right.\right) = 1 - \prod_{t=1}^T \; \left[ \,1 - q_t\left( i \left| \alpha \right. \right)\, \right] \;\;.
\label{eq:visited} 
\end{equation} 
The term $1-q_t\left(i \left| \alpha \right. \right)$ counts the probability
that the player has not placed a bid on value $i$ at stage $t$. The probability
that the player has not bid on value $i$ at any stage is therefore
the product of this single step probabilities. Finally, the probability
that the player has placed a bid on value $i$ at least once
is calculated as the probability to have bid on value $i$ an arbitrary 
number of times minus the probability to have never bid on value $i$.
Notice that the model assumes that players have no memory because 
they are allowed to bid on the same value more than once.
This is, however, unlikely to happen in the L\'evy 
and ballistic regimes (i.e., $\alpha < 3$). Also,
as in the case of real auctions, if the same player
bids more than once on the same value this 
fact does not invalidate the uniqueness of the bid which
is still considered as unique unless another player
places a bid on that value. 
\\
The probability that a generic player, performing
$T$ total bids, has placed a bid on value $i$ is then
\begin{equation} 
p_T\left(i\right) = \int_{\alpha_1}^{\alpha_2} d\alpha \; s_T\left(i \left| \alpha \right. \right) \, g\left(\alpha\right) 
%= \int_{\alpha_1}^{\alpha_2} d\alpha \;  s\left(i, \alpha, T\right) 
\label{eq:final_multi}  
\end{equation} 
and can be used in place of the one appearing 
in Eq.(\ref{eq:single_gen}) in order to
calculate the remaining quantities $u_T\left(i \right)$, $l_T\left(i\right)$ and $w_T\left(\alpha\right)$ by using 
Eqs.~(\ref{eq:unic}),~(\ref{eq:unic_low}) 
and~(\ref{eq:prob_win}), respectively.

\subsection{Evolutionary Model}
\noindent In order to understand how an optimal 
strategy can become shared across individuals,
it is natural to adopt an evolutionary 
framework~\cite{nowak}. In this respect, our model can be implemented in terms
of competing individuals that are selected on the basis of their success
in the searching process. In the spirit of the fundamental 
Moran process~\cite{moran},
at the end of each game, the winner of the
auction generates an offspring to which
transmits her/his search exponent  $\alpha$. The new individual enters 
the population endowed with an exponent $\alpha + \xi$ 
(with $\xi$ random mutation), while a randomly extracted individual 
is removed in order to maintain the
population size constant. Basically, the pdf
of the winning exponents of the former generation
corresponds to the fitness function of the evolutionary model.  

\subsubsection{Absence of mutations}
\noindent Let us first consider the case in which
losers copy the strategy of the winners without
errors. Imagine to have $N$ players at each generation.
They play the game by performing $T$ bids each.
Denote with $e$ the number of the generation.
At the beginning, we set $e=1$. 
Then we follow the scheme:
%%%%%%%%%%%%%%%%%%%%
\begin{enumerate}
\item{Players randomly pick 
strategies $\alpha$ from the pdf 
$g^{(e)}\left(\alpha\right)$;}
\item{They play the game. The result is the
pdf $w_T^{(e)}\left(\alpha\right)$, which quantifies
the probability that $\alpha$ was a winning strategy;}
\item{Set $g^{(e+1)}\left(\alpha\right) = w_T^{(e)}\left(\alpha\right)$,
increment $e \to e+1$, and go back to point 1}.
\end{enumerate}
%%%%%%%%%%%%%%%%%%
\noindent The former procedure describes
the evolution of a population of players 
under repeated games.
Setting $g^{(e+1)}\left(\alpha\right) = w_T^{(e)}\left(\alpha\right)$ ensures
that the players of the new generation have the tendency to pick
winning strategies instead of losing ones.
This can be better understood by writing the master
equation
%\begin{widetext}
\begin{equation}
\begin{array}{l}
g^{(e+1)}\left(\alpha\right) - g^{(e)}\left(\alpha\right) = 
\\
w_T^{(e)}\left(\alpha\right) \left[ 1 - g^{(e)}\left(\alpha\right) \right] - \left[ 1 - w_T^{(e)}\left(\alpha\right) \right] g^{(e)}\left(\alpha\right) \;\; ,
\label{eq:master}
\end{array}
\end{equation}
%\end{widetext}
from which one can easily
obtain $g^{(e+1)}\left(\alpha\right) = w_T^{(e)}\left(\alpha\right)$.
Eq.~(\ref{eq:master}) tells us that the variation
in the population of players with exponent
$\alpha$ increases as the probability
that a player with strategy $\alpha$ wins [$w_T^{(e)}\left(\alpha\right)$]
times the probability
to have other players with strategies different from $\alpha$ 
[$ 1 - g^{(e)}\left(\alpha\right)$], and
decreases as the probability that
a player with strategy different from $\alpha$
wins [$1 - w_T^{(e)}\left(\alpha\right)$] times the probability to 
find a player with strategy equal to $\alpha$ [$g^{(e)}\left(\alpha\right)$]. 
In other words, 
the probability of reproduction
of a player is proportional to the ability of player
to win the game.
The evolution rules resemble a Moran process where
selection is made according to a fitness function
here defined as the probability to win the game~\cite{moran}.

\subsubsection{Random mutations}
\noindent A more natural assumption is to 
formulate a model where, each time a losing player changes
strategy, she/he copies the exponent of the winner
plus some random variation. Assume that the variation $\xi$
is randomly extracted from a pdf
$y\left(\xi, \alpha, \vec{\mu}\right)$,
explicitly  dependent on $\alpha$ and a
set of parameters $\vec{\mu}$. The
master equation describing the evolution becomes
\[
\begin{array}{l}
g^{(e+1)}\left(\alpha\right) - g^{(e)}\left(\alpha\right) =\\ \\ 
+  \int d\beta \int d\xi \; \delta\left(\alpha - \beta + \xi\right) \; y\left(\xi, \beta, \vec{\mu}\right) \times
\\  w_T^{(e)}\left(\beta\right) \left[ 1 - g^{(e)}\left(\beta\right) \right]
\\
\\
 - \left[ 1 - w_T^{(e)}\left(\alpha\right) \right] g^{(e)}\left(\alpha\right)
\end{array}  
\;\; ,
\]
with $\delta \left( x \right) = 1$ if and only if $x=0$, 
while  $\delta \left( x \right) = 0$ otherwise. 
The gain term stands for the probability that the generic exponent $\beta$
represents the winning strategy [$w_T^{(e)}\left(\beta\right)$]
times the probability to have other players with exponent
different from $\beta$ [$ 1 - g^{(e)}\left(\beta\right)$]. This term
is then convoluted with the noise over all possible values of $\beta$:
this quantifies the probability
that a random mutation changes the exponent from $\beta$ to $\alpha$
[$\delta\left(\alpha - \beta + \xi\right)$].
The loss term is simply the probability that $\alpha$ is not the winning 
strategy [$1 - w_T^{(e)}\left(\alpha\right)$] 
and therefore players  with exponent $\alpha$ moves
away from it at rate $g^{(e)}\left(\alpha\right)$. The delta function
imposes the condition $\alpha - \beta + \xi = 0$ and 
the former equation 
reduces to
\begin{equation}
\begin{array}{l}
g^{(e+1)}\left(\alpha\right) = 
\\
+ w_T^{(e)}\left(\alpha\right) \, g^{(e)}\left(\alpha\right) 
\\
+  \int d\beta \; y\left(\beta - \alpha, \beta, \vec{\mu} \right) \; w_T^{(e)}\left(\beta\right) \left[ 1 - g^{(e)}\left(\beta\right) \right]
\end{array}
 \;\; .
\label{eq:master_noise}
\end{equation}
The pdf of the adopted strategies  converges to stability 
whenever exists $e$
for which $g^{(e+1)}\left(\alpha\right) = g^{(e)}\left(\alpha\right)$,
which automatically implies that $ w_T^{(e+1)}\left(\alpha\right) = w_T^{(e)}\left(\alpha\right)$.

%%%%%%%%%%%%%%%%%
\begin{figure}[!t]
\begin{center}
\includegraphics[width=0.45\textwidth]{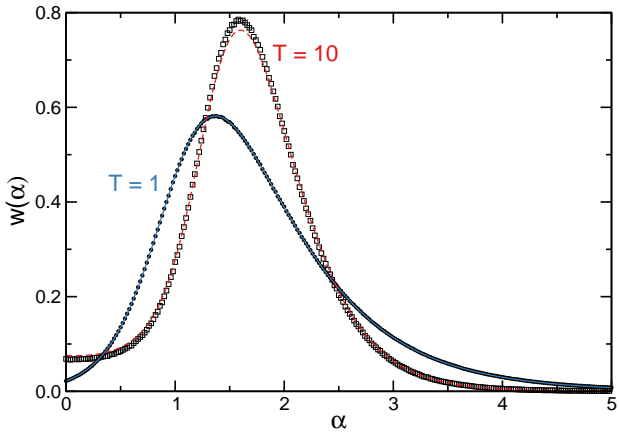}
\caption{(Color online) Probability density function
of the winning strategy $w\left(\alpha\right)$
in the case in which players' strategies are
randomly extracted from a uniform probability density
function $g\left( \alpha \right)$ defined
over the interval $\left(0,5 \right)$.
We set $N=100$ and $M=1000$ (parameter values
similar to those observed in real 
auctions~\cite{radicchi11}), and consider for simplicity
only the case of a single bid $T=1$ 
and the case in which all players perform $T=10$ bids.
We show in both cases the comparison between the results obtained with
numerical simulations ($T=1$ filled black circles, $T=10$ empty black squares) 
and those obtained with
the numerical integrations of the various equations 
[$T=1$ full (lower) line,
 $T=10$  dashed (upper) line]. Note that the optimal value 
of $\alpha$ depends on $T$
and also the parameters values $N$ and $M$. However, this 
is a weak dependence and the peak value of 
probability density function
of the winning strategy is in the range $1.2$ to $1.7$ for a wide
range of possible values of $N$ and $M$ and 
a moderately broad range of values of $T$.
}
\label{fig:T1_l0}
\end{center}
\end{figure}
%%%%%%%%%%%%%%%%%

%%%%%%%%%%%%%%%%%
\begin{figure*}[!htb]
\begin{center}
\includegraphics[width=0.9\textwidth]{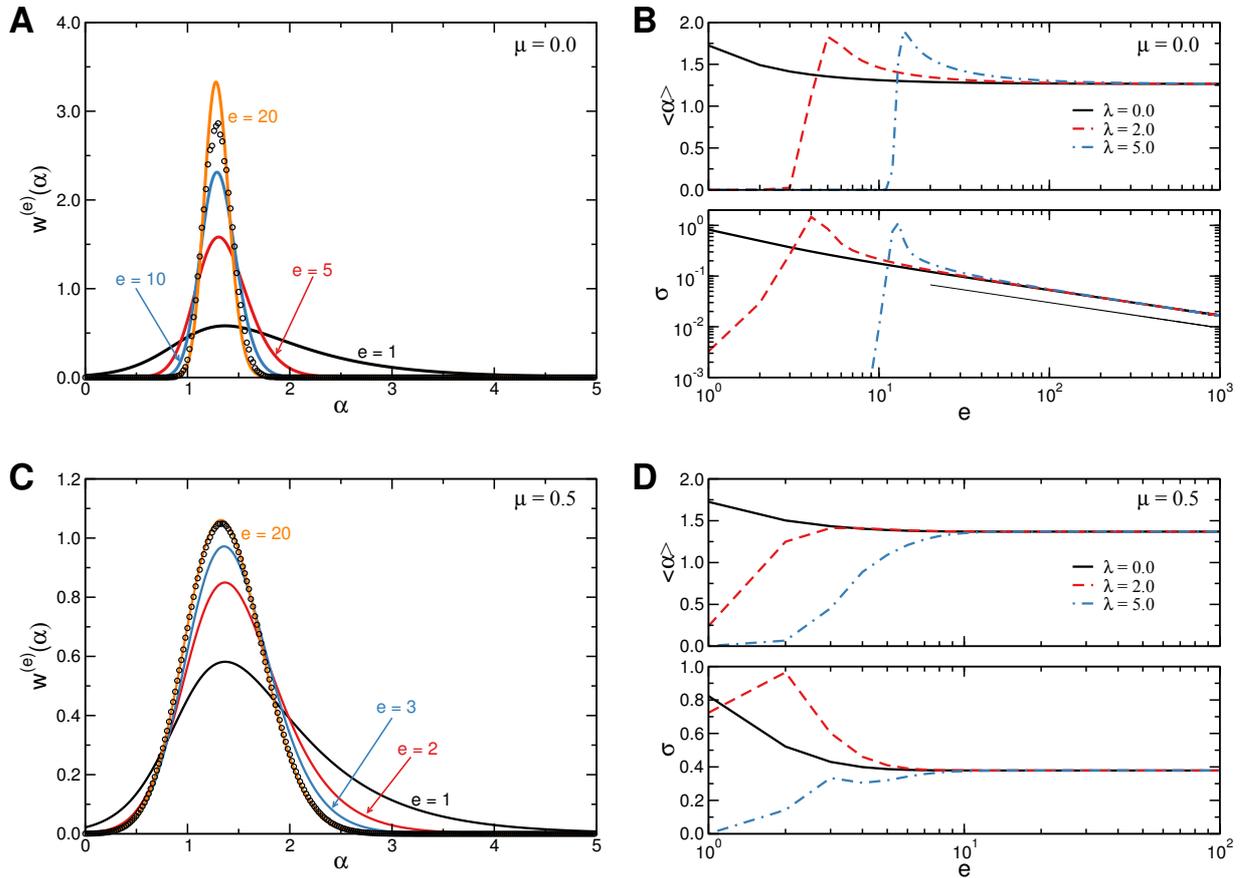}
\caption{(Color online) Probability distribution 
function $w^{(e)}\left(\alpha\right)$
of the winning strategies at generation $e$. Here we set
$M=1000$, $N=100$ and $T=1$.
The mutation $\xi$ is randomly extracted from
a uniform distribution centered in zero and width $2\mu$:
we consider the cases $\mu=0.0$ (panels A and B, 
absence of mutation) and $\mu=0.5$ 
(panels C and D).
In A and C, the starting distribution for the exponents is
$g^{(1)}\left(\alpha\right)=1/5$ 
if $\alpha \in \left(0,5\right)$ and  zero otherwise.
Already after one generation 
($e=1$, lower black line), the peak of the distribution 
is around $\alpha\simeq 1.5$. As the number of generations
increases, the distribution becomes more and even more
peaked around a specific value of $\alpha$,
and reaches a stationary distribution. 
The asymptotic distribution has finite width
for $\mu>0$, while is a delta function
for $\mu=0$. As a term of comparison,
we show also the results obtained 
with numerical after $e=20$ generations (black circles).
In B and D, we consider
initial distributions for the exponents 
of the type $g^{(1)}\left(\alpha\right) \sim \alpha^{-\lambda}$ if
$\alpha \in \left(0,5\right)$ and  $g^{(1)}\left(\alpha\right)=0$ otherwise.
We show the results for the three cases: 
$\lambda=0$ (uniform, black full line), 
$\lambda=2$ (power-law, red dashed line) and
$\lambda=5$ (exponential, blue dot-dashed line). The asymptotic
distribution does not depend on the initial distribution
and the peak value of the
asymptotic probability density function does not
depend on $\mu$. Independently of the value of
$\lambda$, as the number of generations $e$ increases,
the average value $\langle \alpha \rangle$ and 
the standard deviation $\sigma$
approach the same stationary values: $\langle \alpha \rangle \simeq 1.3$
for any $\mu$, while $\sigma = 0$ (as $1/\sqrt{e}$, 
thin black line in panel B) for $\mu=0.0$
and $\sigma \simeq 0.4$ for $\mu=0.5$.}
\label{fig:evo_T1_l0}
\end{center}
\end{figure*}
%%%%%%%%%%%%%%%%%

%%%%%%%%%%%%%%%%%
\begin{figure*}[!htb]
\begin{center}
\includegraphics[width=0.9\textwidth]{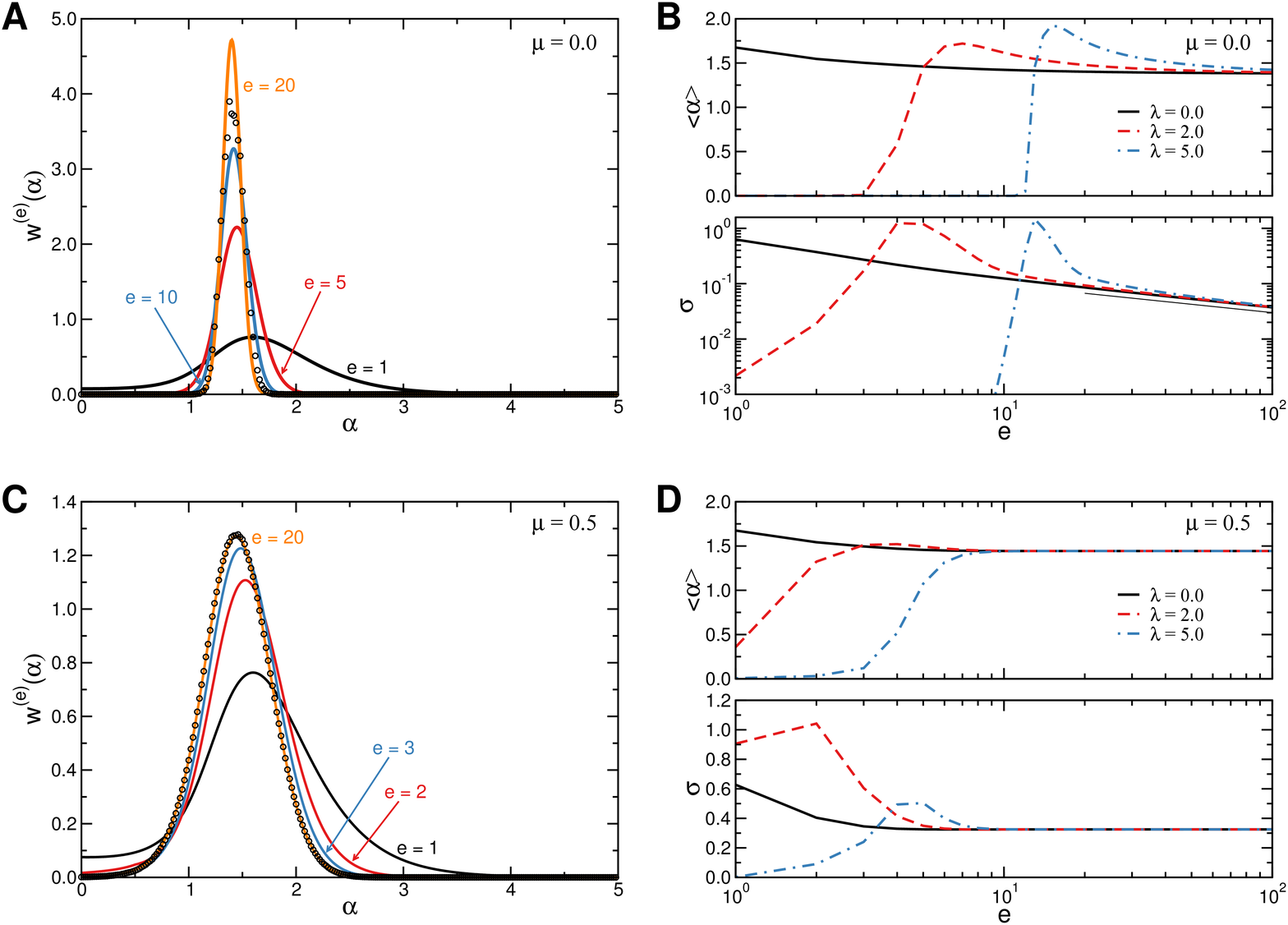}
\caption{(Color online) Same as figure~\ref{fig:evo_T1_l0}, but
with different parameter values. Here we set
$M=1000$, $N=100$ and $T=10$, and consider 
$\mu=0.0$ in panels A and B, while
$\mu=0.5$ in panels C and D. The asymptotic pdf
of the winning strategies have average values $\langle \alpha \rangle \simeq 1.5$ for any value of $\mu$. The standard deviation is $\sigma =0 $ (as $1/\sqrt{e}$) for $\mu=0.0$, and  $\sigma=0.3$ for $\mu=0.5$. For any value
of $\mu$, the best strategy (i.e., the peak of the pdf)
is placed at $\alpha^* \simeq 1.5$.}
\label{fig:evo_T10_l0}
\end{center}
\end{figure*}
%%%%%%%%%%%%%%%%%

\section{Numerical Simulations}
\noindent The former analytical formulation of the model does not allow
to obtain explicit expressions regarding the distributions
of the winning strategies. The various equations can, in fact, be 
only numerically integrated to 
provide a solution of the model. Moreover,
some of the equations contain approximations, and
it is therefore worth asking whether the solutions
obtained with the numerical integration of the equations
are compatible with those obtained by directly simulating the model.
\\
Simulating our model is straightforward.
In each simulation, we use the following
scheme:  
\begin{enumerate}
\item{Extract the exponent values of each of the $N$
players from the given pdf $g\left(\alpha\right)$;}
\item{Simulate the game: for each of the $N$ players, 
extract $T$ integer bid values from the corresponding
power-law distribution, and determine the winner of the auction
(the player who made the LUB) on the basis of these extractions. }
\end{enumerate}
In the case of the evolutionary game model,
at the end of the game we need to change the exponent 
value of one of the losing players, by copying (with or without random
mutations) the strategy of the winner. We then repeat the game. 
A generation corresponds to $N$ exponent changes.
The pdf of the winning strategies of each generation
is computed by repeating the entire procedure many times.

\section{Computational Complexity}
\noindent One could argue why
use a complicated and approximated set of equations
instead of simple and straightforward numerical simulations.
The reason is that the computational time required for
the numerical integration of the model's equations is much much lower
than the one needed for obtaining good estimates
with numerical simulations.
For clarity, we provide here an estimation of the computational complexity
required in both approaches to the solution
of the model.
\\
Consider first the case
$T=1$ (i.e., players make a single bid).
In the case of numerical
simulations with fixed values of the exponents, the time required
to simulate the game grows as
$M + N\, \log{\left(M\right)} \, G$.
 $M$ is the
number of possible bid values and indicates also 
the computational time required to calculate
the transition matrix from the starting position
in the bid space
(i.e., the origin of the lattice) 
to all $M$ possible bid values. 
$N$ is the number of players.
The bid value on which each player
places a bid can be calculated in a time
that scales as $\log{\left(M\right)}$.
Finally $G$ is the 
number of times that we need to simulate the 
same auction model
in order to obtain a good estimation of the 
pdf $w\left(\alpha\right)$.
The computational complexity of the numerical solutions of
equations is $M$, since the computation of 
Eqs.~(\ref{eq:single}), (\ref{eq:single_gen}), (\ref{eq:unic}), (\ref{eq:unic_low}) and
(\ref{eq:prob_win}) require a computational time
that grows as $M$.
\\
For general values of $T$, the computational complexity
of numerical simulations is  
simply incremented by a factor $T$
and grows therefore as $T\, M + T\, N\, \log{\left(M\right)} \, G$.
The time required for the numerical integration
of the equation differently grows as $T\, M^2$.
The most computationally expensive
calculation is the one of Eq.~(\ref{eq:ran_walker}) that requires
a time growing as $M^2$, and this computation
has to be repeated $T$ times.
%The numerical solution of the equations is therefore
%much faster, and, in spite
%of the approximations, provides results consistent with those
%obtained with ``computationally expensive'' numerical simulations.

\section{Results}
\noindent In all results, we consider the values $N=100$ and $M=1000$.
The choice of these parameter values is justified because 
they are of the same order of magnitude as those measured
in real auctions~\cite{radicchi11}.
In Fig.~\ref{fig:T1_l0}, we plot the 
pdf $w\left(\alpha\right)$ for $T=1$ and $T=10$.
In both cases, $g\left(\alpha\right)=1/5$ if
$\alpha \in \left(0,5\right)$, while  $g\left(\alpha\right)=0$
otherwise. Players randomly choose strategies that 
correspond to ballistic motion ($\alpha \leq 1$),
diffusive motion ($\alpha>3$) and super-diffusive
motion or L\'evy flight ($1 < \alpha \leq 3$).
Assigning a flat initial distribution corresponds
to assuming that players don't know which strategy is
better for winning the game and all strategies
are therefore {\it a priori} equivalent.
It is interesting to see that already in this situation there is
a clear advantage for players that perform L\'evy flights 
and whose exponents are in the range $1.2-1.5$.
It is also worth noting that the solutions obtained
with the numerical integration of the equations
are perfectly consistent with the results obtained with
numerical simulations.

\noindent More interesting is the case of evolutionary games.
For simplicity, here we consider the
case in which the distribution of the random mutations
is a rectangular window of width $2\mu$: 
if $\alpha$ is the winning strategy, the new exponent
is equal to a random number taken from the 
uniform distribution into the interval $\left(\alpha-\mu,\alpha+\mu\right)$.
In the case in which the possible values of the
exponents are bounded
in the interval $\left(\alpha_1, \alpha_2\right)$, we have to include the effect
of the boundaries and write
\begin{equation}
y \left(\xi, \alpha, \alpha_1, \alpha_2, \mu \right) =\left\{
\begin{array}{ll} \left[ \ell_1 + \ell_2 \right]^{-1} & \textrm{ , if } \;  \xi \in  \left( -  \ell_1 , \ell_2  \right)
\\
0 & \textrm{ , otherwise}
\end{array}
\right.
 \;,
\label{eq:noise}
\end{equation}
where $\ell_1 = \min{ \left( \alpha - \alpha_1 , \mu \right) }$ and $ \ell_2 =  \min{ \left( \alpha_2 - \alpha , \mu \right) }$.
We show in Fig.~\ref{fig:evo_T1_l0} the results valid for $T=1$
and in Fig.~\ref{fig:evo_T10_l0} those obtained for $T=10$.
These figures show results
that are slightly different, but qualitatively identical.
The asymptotic pdf $w^{(\infty)}\left(\alpha\right)$
of the winning strategies does not depend
on the initial distribution $g^{(1)}\left(\alpha\right)$.
In absence of mutation, the limiting pdf is a delta function
centered around an optimal strategy $\alpha^*$.
The convergence to such pdf is only asymptotic, since
the width of the distribution goes to
zero as $1/\sqrt{e}$ and is therefore finite at any finite 
generation $e$. If mutations
are allowed (i.e., $\mu>0$), 
$w^{(\infty)}\left(\alpha\right)$
is reached after a finite number of iterations.
The asymptotic distribution has a finite width.
The number of iterations required to reach stability
depend on the initial pdf and the mutation rate $\mu$,
while its shape only on $\mu$. In particular, the peak of the pdf
is still at $\alpha^*$, the same value as the one measured
in absence of mutations.
The value of $\alpha^*$ depends on the parameters of the model
$N$, $M$ and $T$ but, for values consistent with those of
real auctions ($N=50$ to $N=200$, $M=500$ to $M=10000$
and $T=1$ to $T=100$), $\alpha^*$ ranges from $1.2$ to $1.7$.

\section{Conclusions} 
\noindent A large wealth of empirical evidences suggest that
animals have a specific, and apparently innate, strategy to search 
an unknown physical environment~\cite{viswanathan08}.
The distances  between two consecutive positions
in the space are distributed according to
a power-law probability density function.
More interestingly, the exponent of the power-law
distribution is close to the one that guarantees
the most efficient search in an environment with scarce resources.
Evolutionary considerations account for the high
efficiency of the searching strategies of animals,
since in a competitive environment only the
fittest are able to survive and reproduce.

\noindent A similar behavior has been observed also in how
human players explore the bid space in on-line auctions~\cite{radicchi11}.
In this case, the environment is not physical but
abstract. Nevertheless players adopt searching
strategies for the winning bid value that are close to
optimality: bid change amounts are
power-law distributed, and the exponents of the
power-laws are close to the value
that can guarantee the highest winning chances. 

\noindent In this paper, we have provided a novel
interpretation of this empirical evidence and used
evolutionary considerations to explain 
the optimality of the observed exponents.
We have introduced a Moran-like model in which
the reproductive fitness of the players is proportional to their success
in searches~\cite{moran}. The player winning the auction reproduces,
in the sense that the strategy of the winner is transmitted to another 
randomly extracted individual.  We have considered both the cases 
of error-free reproduction and of transmission with mutation. 
We have described the model analytically through a set of equations
whose numerical solution is in excellent agreement with direct
agent-based simulations. We have shown that the model is extremely 
robust with respect to the choice of the different parameters, producing
results in good agreement with the ones observed in the empirical data.

\

\noindent In summary, looking at activity patterns in the web~\cite{lazer},
our previous work  \cite{radicchi11} suggested  that humans and other animals share the same,
apparently innate, strategy to search 
in an unknown, physical or mental, environment. 
Here we have shown that an evolutionary approach allows to account 
for the optimality of the observed exponents, in agreement
with the view according to which the ability 
to understand and  be effective
in the natural world is likely to be innate~\cite{ninduction}. 
This is the case for example of locomotion and 
perceptual-motor control~\cite{alexander}, hunting and foraging~\cite{stephens} 
or nest building~\cite{healy}. We have provided a new example with the 
remarkable novelty that it concerns a mental search process. 
While it  is well known that humans share the intuition 
that numbers map into space~\cite{dehaene08}, our work
indicates that they might have developed an innate knowledge 
about the best way to move in it.

\begin{acknowledgments}
\noindent The authors are indebted to Lu\'{\i}s A. N. Amaral for helpful discussions on the subject of this article. F. Radicchi acknowledges support from the Spanish Ministerio de
  Ciencia e Innovaci\'{o}n through the Ram\'on y Cajal program.
\end{acknowledgments}

\end{document}